\documentclass{optica-article}
\journal{opticajournal} 

\articletype{Research Article}

\usepackage{lineno}
\usepackage{adjustbox}

\begin{document}

\title{Generation of photon pairs through spontaneous four-wave mixing in thin nonlinear layers}

\author{Changjin Son,\authormark{1,2,*} Samuel Peana,\authormark{3} Owen Matthiessen,\authormark{3} Artem Kryvobok,\authormark{3} Alexander Senichev,\authormark{3} Alexandra Boltasseva,\authormark{3} Vladimir M. Shalaev,\authormark{3} and Maria Chekhova\authormark{1,2}}

\address{\authormark{1}Max Planck Institute for the Science of Light, 91058 Erlangen, Germany\\
\authormark{2}Friedrich-Alexander Universität Erlangen-Nürnberg, 91058 Erlangen, Germany\\
\authormark{3}School of Electrical \& Computer Engineering, Birck Nanotechnology Center, Purdue University, 1205 West State Street,
West Lafayette, Indiana 47907, USA}

\email{\authormark{*}changjin.son@mpl.mpg.de} 


\begin{abstract*} 
Pairs of entangled photons are crucial for photonic quantum technologies. 
The demand for integrability and multi-functionality suggests ‘flat’ platforms - ultrathin layers and metasurfaces - as sources of photon pairs. 
Despite the success in the demonstration of spontaneous parametric down-conversion (SPDC) from such  sources, there are almost no works on spontaneous four-wave mixing (SFWM) – an alternative process to generate photon pairs. Meanwhile, SFWM can be implemented in any nanostructures, including ones made of isotropic materials, which are easier to fabricate than crystalline SPDC sources. 
Here, we investigate photon pair generation through SFWM in subwavelength films of amorphous silicon nitride (SiN) with varying nitrogen content. For all samples, we demonstrate two-photon quantum correlations, indicated by the normalized second-order correlation function $g^{(2)}(0)$: it exceeds $2$ and decays as the pump power increases.
By observing two-photon interference between SFWM from the SiN films and the fused silica substrate, we find the third-order susceptibilities of films with different nitrogen content. 

\end{abstract*}

\section{Introduction}
Pairs of entangled photons are of great importance for quantum technologies, including quantum communication, computation, and sensing. 
The usually employed means to generate such pairs are spontaneous parametric down-conversion (SPDC) and spontaneous four-wave mixing (SFWM), utilizing nonlinear susceptibilities of the second ($\chi^{(2)}$) and third ($\chi^{(3)}$) order, respectively. 
The need for more compact sources of photon
pairs led to the implementation of SPDC on ‘flat’ platforms, such as ultrathin nonlinear layers \cite{santiago2021entangled}, including those which are only few-atomic layers thick \cite{guo2023ultrathin}, nanoantennas \cite{marino2019spontaneous}, and resonant metasurfaces \cite{santiago2021photon}. Importantly, the relaxed phase-matching condition in such sources \cite{okoth2019microscale} simplifies quantum state engineering. 
Such sources can be made of highly nonlinear materials~\cite{guo2023ultrathin, kallioniemi2024van}, provide tunable polarization \cite{sultanov2022flat, ma2023polarization, ma2024generation, weissflog2024tunable} or spatial \cite{zhang2022spatially} entanglement, and enable several entanglement links \cite{santiago2022resonant} or  directions of emission \cite{son2023photon,weissflog2024directionally} simultaneously. 
Pairs generated in thin layers feature very broad spectra in both angle and frequency as well as large degrees of continuous-variable entanglement~\cite{okoth2019microscale}. 
Resonances in  metasurfaces enhance the rate of pair emission by several orders of magnitude~\cite{santiago2021photon, ma2024generation, santiago2022resonant}, in the vicinity of resonance wavelengths.

All of the works mentioned above employ  SPDC for the generation of photon pairs. 
Meanwhile, there are almost no experiments where entangled photons are obtained through SFWM in nanoscale sources, although this third-order process is widely used to obtain entangled photons from fibers \cite{sharping2001four}, waveguides \cite{fukuda2005four}, and microresonators \cite{kippenberg2004kerr, wang2024progress}. Despite theoretical studies \cite{xu2022enhanced, yuan2019spatiotemporally}, and experiments with seeded FWM \cite{nielsen2017giant, yang2022stimulated, caspani2016enhanced, carnemolla2021visible}, to the best of our knowledge, there is only a single paper on SFWM in a ‘flat’ platform, namely a 100 nm-thick layer of carbon nanotubes \cite{lee2017photon}. 
However, the relatively high absorbance of this material at telecom wavelengths \cite{nishihara2022empirical,song2020giant} limits its applications. 

Here, we demonstrate SFWM in subwavelength films of silicon nitride (SiN) at various nitrogen concentrations. 
A higher N content reduces $\chi^{(3)}$ \cite{ning2013third,tan2018nonlinear,ding2019third} and thus the rate of SFWM, but at the same time can suppress photoluminescence (PL), which is especially detrimental for nanoscale sources of photon pairs \cite{sultanov2023temporally}, and increases the damage threshold.
Additionally, with higher silicon content SiN can possess similar nonlinearity and maintain larger band gaps compared to layers of pure silicon~\cite{ooi2017pushing}.

In SFWM, two pump photons at frequency $\omega_p$ convert into a pair of photons at frequencies $\omega_s$ (signal) and $\omega_i$ (idler), so that the energy is conserved, $\omega_s+\omega_i=2\omega_p$. 
The momentum conservation requires a zero mismatch $\Delta k=k_s+k_i-2k_p$ for the signal, idler, and pump wavevectors $k_{s,i,p}$. Although $\Delta k=0$ is usually called the phase matching condition, the {\it phase mismatch} between the signal-idler pair and the two pump photons is $\Delta\phi=l\Delta k$, where $l$ is the nonlinear sample thickness. This phase mismatch is small if $l$ is below the coherence length $L_{coh} = \pi/\Delta k$. This is why for thin films, SFWM is expected to be phase matched automatically over a very broad spectral range.

Accordingly, our subwavelength SiN samples generate signal and idler photons with wavelengths up to an octave apart. Moreover, the substrates on which they are coated also generate phase-matched broadband SFWM: although they are thicker, their $L_{coh}$ is longer.
This leads to the two-photon interference between the pairs generated from the samples and from the substrates. 
From this interference, we infer the contribution of the SiN films to the SFWM signal and eventually their $\chi^{(3)}$ values.

\section{SFWM in subwavelength SiN layers}
We fabricated SiN samples using high-density plasma chemical vapor deposition (HDPCVD, see Supplement 1 Sec. 1 - 3). Films of different thicknesses $l$ were all coated on 500 $\mu$m fused silica substrates and had different concentrations of N, quantified by the growth precursor flow ratio $x$ of $\mathrm{N_2}$ and $\mathrm{SiH_4}$ gases.
Table 1 summarizes the properties of our samples: ratio $x$, thickness $l$, refractive indices at the wavelengths used in the first series of measurements ($1030$ nm, $1550$ nm and $770$ nm for the pump, idler, and signal photons, respectively), the corresponding nonlinear coherence length $L_{coh}$, and $\chi^{(3)}$ evaluated from third harmonic generation (THG); see Supplement 1 Sec. 5 for more details. For films A and B, the most silicon rich samples, the value of $\chi^{(3)}$ is close to that reported for silicon ($2.8\times10^{-18}~\mathrm{m^2/V^2}$)~\cite{boyd2020nonlinear} as expected.  
We see that $\chi^{(3)}$ continuously reduces with an increase in N concentration, as expected, because N broadens the band gap and therefore brings the excitation further from the resonance. 
The level of the PL also strongly depends on the N concentration and can be dramatically suppressed at different concentrations.
In the bottom row, we also show the parameters of the fused silica substrate.

\begin{table}[h]
    \centering
    \begin{adjustbox}{width=0.8\columnwidth}
    \begin{tabular}{c c c c c c c}
    \hline
     Sample ($x$) &$l$ ($\mu$m) & $n_p$ & $n_{IR}$ & $n_{VIS}$ & $L_{coh}$ ($\mu$m) & $\chi_{THG}^{(3)}$ (m$^2$/V$^2$)
     \\
     \hline
     A (0.18) & 0.818 & 2.93 & 2.87 & 3.04 & 4.92  & 2.3$\times10^{-18}$ \\
     B (0.37) & 0.511 & 2.63 & 2.58 & 2.71 & 6.32  & 1.5$\times10^{-18}$ \\
     C (0.56) & 0.697 & 2.33 & 2.31 & 2.39 & 9.11 & 2.1$\times10^{-19}$ \\
     D (0.75) & 0.457 & 2.18 & 2.17 & 2.21 & 19.3 & 6.8$\times10^{-20}$ \\
      Substrate & 500 & 1.4501 & 1.4440 & 1.4539 & 486 & - \\
     \hline
    \end{tabular}
    \end{adjustbox}
    \caption{SiN samples used in the experiment, with their ratios $x$, thicknesses $l$, refractive indices at the pump, idler (IR) and signal (visible) wavelengths (1030 nm, 1550 nm, and 770 nm, respectively), corresponding nonlinear coherence lengths $L_{coh}$,  and the $\chi^{(3)}$ measured from THG. The last row shows the parameters of the fused silica substrate.}
    \label{tab:samples}
\end{table}

\begin{figure}[h]
    \centering
    \includegraphics[width=0.5\linewidth]{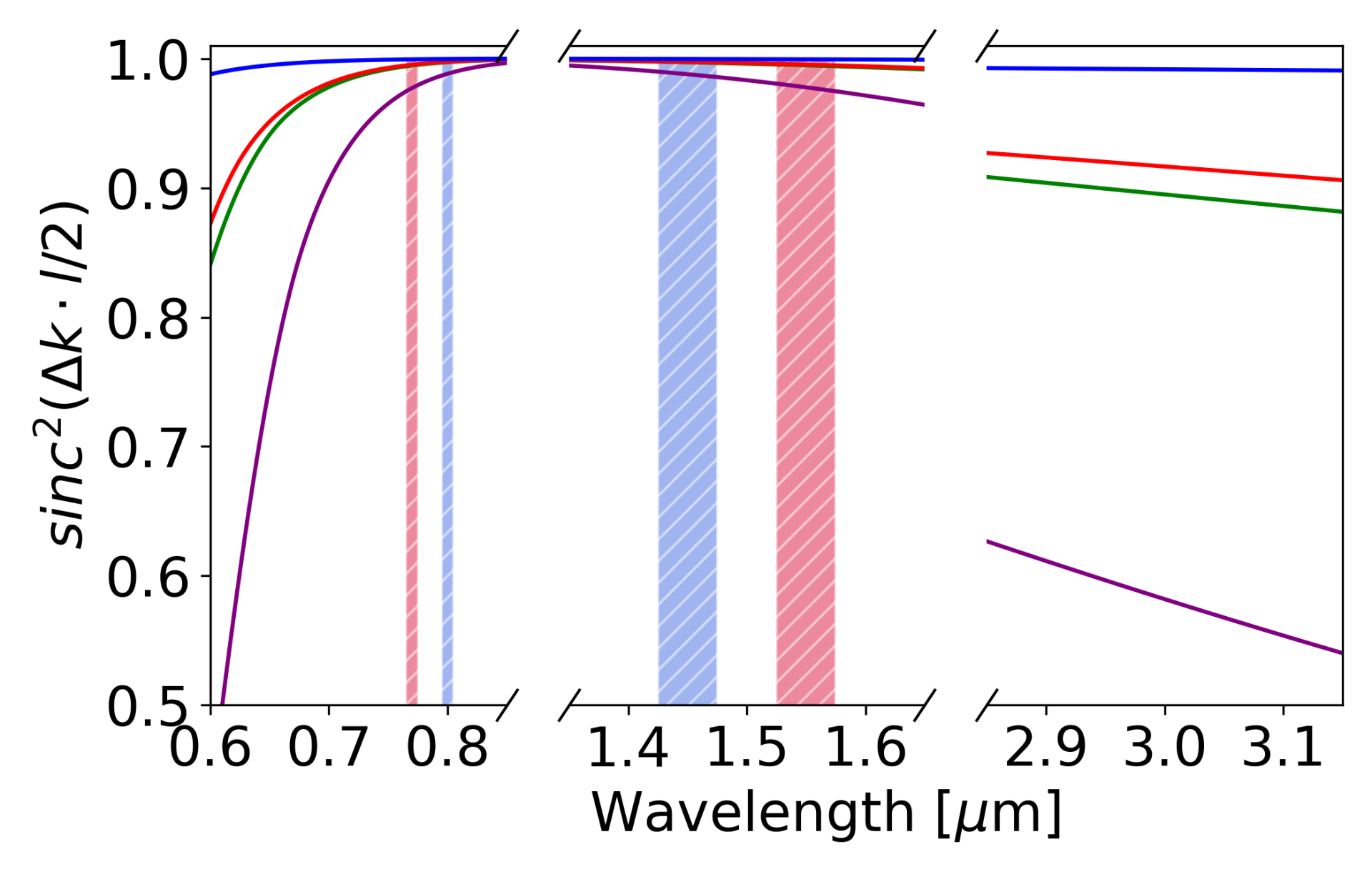}
    \caption{The phase matching function calculated vs the output photon wavelength for samples A (purple), B (green), C (red), and D (blue).
    The blue and red shaded areas indicate the detection band sets used in experiment.}
    \label{fig:sinc function}
\end{figure}

Figure~ \ref{fig:sinc function} shows the phase matching function $\hbox{sinc}^2(\Delta k l/2)$, which is the main factor determining the SFWM spectrum, for all samples under pumping at $1030$ nm. The spectral width reduces as the Si content increases, but still exceeds two octaves for all samples. Out of this broad spectrum, in our experiments we filtered separate bands for the signal and idler photons using available bandpass filters, but roughly satisfying the energy conservation: [$1550$ nm, $770$ nm] for one experiment and [$1450$ nm, $800$ nm] for the other (shown by red and blue shaded bands). The bandwidths of the filters were $10$ nm for visible and $50$ nm for IR photons.

\begin{figure}[h]
\centering
\includegraphics[width=0.7\columnwidth]{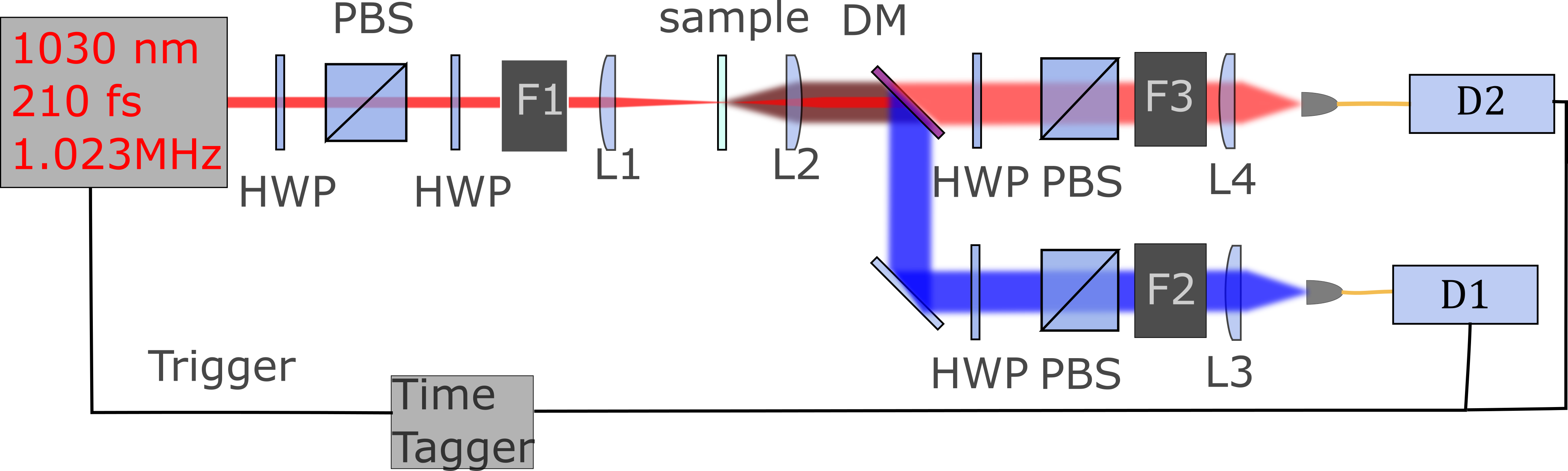}
\caption{Experimental setup: lens L1 focuses the pump on the sample; signal and idler photons are collimated by lens L2, separated by dichroic mirror DM, and coupled into  detectors D1, D2 by lenses L3, L4, respectively. 
Their polarization is analyzed by an HWP and PBS in each arm. 
Filter F1 eliminates unwanted wavelengths in the pump beam and bandpass filters F2, F3 select the signal and idler photon detection bands (marked in Fig.~\ref{fig:sinc function}).
Photon detection events are registered by a time tagger triggered by pulses from the laser.
}
\label{setup}
\end{figure}

In our experiment (Fig.~\ref{setup}), the pump (210 fs pulses at 1030 nm with repetition rate 1.023 MHz), whose power and polarization were varied using two half-wave plates (HWP) and a polarizing beam splitter (PBS), was cleaned with bandpass filter F1 ($10$ nm centered at $1030$ nm) and focused into 100 $\mu$m on the sample by lens L1 (f=100 mm). 
The generated photons were collimated by lens L2 (NA=0.5) and spectrally separated by a long-pass dichroic mirror with the cut-on wavelength 950 nm. 
Signal (visible) and idler (IR) photons were filtered by bandpass filters F2,F3 (Fig.~\ref{fig:sinc function}). Signal and idler photons were then coupled by lenses L3 (f=30 mm) and L4  (f=50 mm) into 100 $\mu$m multimode fibers and sent to single-photon detectors D1 (Perkin\&Elmer, SPCM-AQRH-16-FC) and D2 (IDQ, ID220), respectively. The detection events were registered by the time tagger, triggered by pulses from the laser. 
This measure was aimed at reducing the level of PL, which is a serious obstacle in all   experiments where photon pairs are generated through nonlinear optical effects~\cite{sultanov2023temporally}.

\begin{figure}[h]
\centering
\includegraphics[width=0.7\columnwidth]{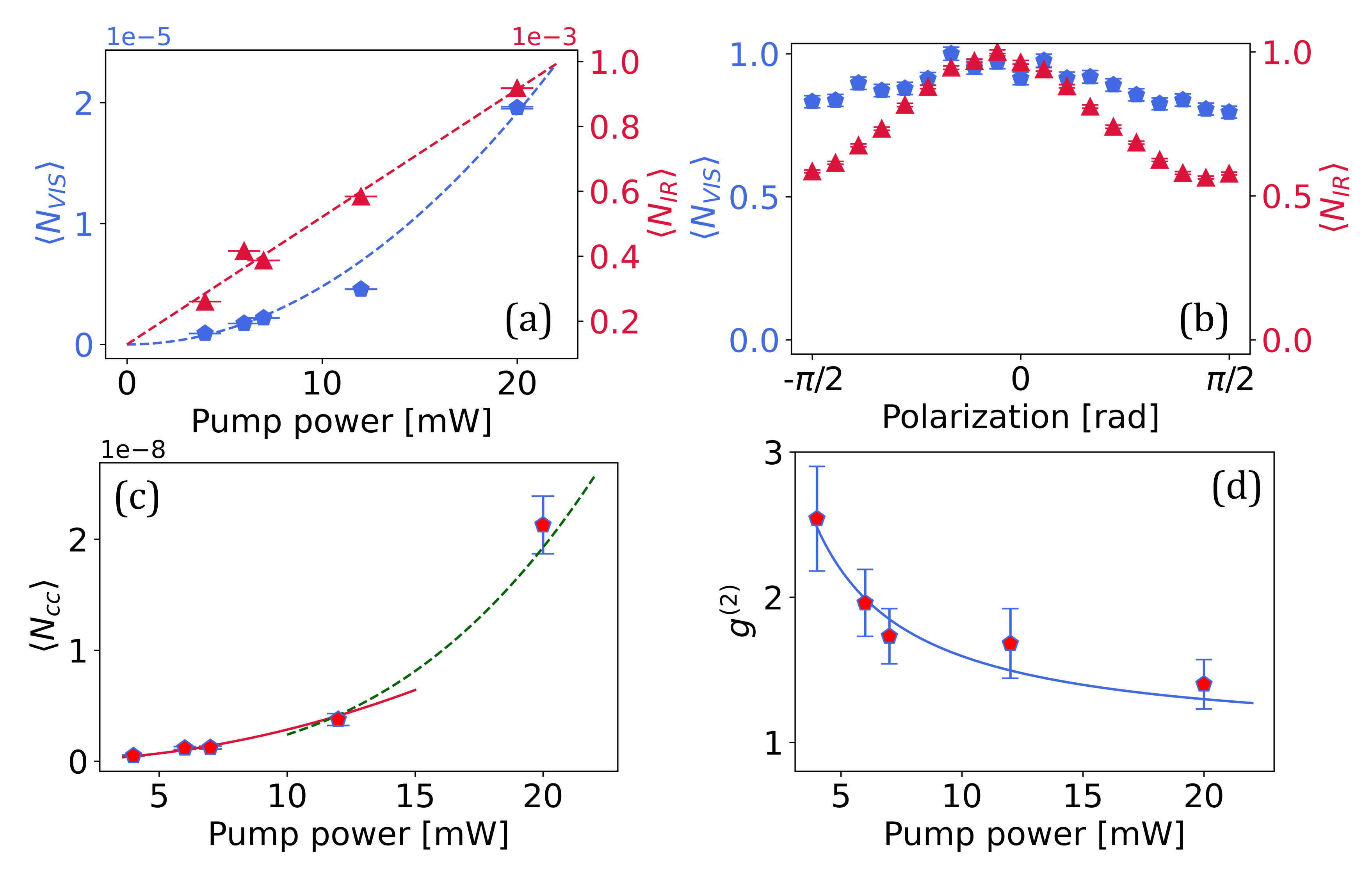}
\caption{SFWM in sample C. 
(a) Mean photon numbers per pulse at 770 nm (blue) and 1550 nm (red) vs the pump power, with their  fits. 
(b) Polarization dependence of the mean photon numbers per pulse at 770 nm (blue) and 1550 nm (red), measured by rotating HWPs in the signal and idler channels, respectively.
 0° corresponds to the pump polarization. 
(c) Mean number of coincidences per pulse vs the pump power, with the quadratic (solid)  and cubic (dashed) fits. 
(d) $g^{(2)}(0)$ vs the pump power.
}
\label{Data_SiN}
\end{figure}

In the first series of measurements, we used the band set $1550$ nm and $770$ nm as F3 and F2, respectively. Figure~\ref{Data_SiN} shows the results of measurements on sample C, all gated with electronic pulses from the laser (similar measurements have been performed for other samples; see Supplement 1 Sec.9 for that).
Despite gating, single-photon counts (panel a) still mostly originate from the PL.
For the IR channel (red), this is clear from the power dependence, which is linear, while the rate of photon emission due to SFWM should depend quadratically on the pump power. 
The rate of photons in the visible channel (blue) indeed has a quadratic dependence on the pump power. 
However, this emission is almost unpolarized (Fig. \ref{Data_SiN} b), while SFWM in an isotropic material such as SiN should be polarized similarly to the pump. 
This indicates that the visible photons are mostly the result of two-photon pumped PL, which explains the quadratic rate dependence on the pump power.
The IR PL is partially polarized due to the small (0.5 eV) frequency shift from the pump \cite{diener1998two} (see Supplement 1 Sec. 7).

\begin{figure}[h]
\centering
\includegraphics[width=0.7\columnwidth]{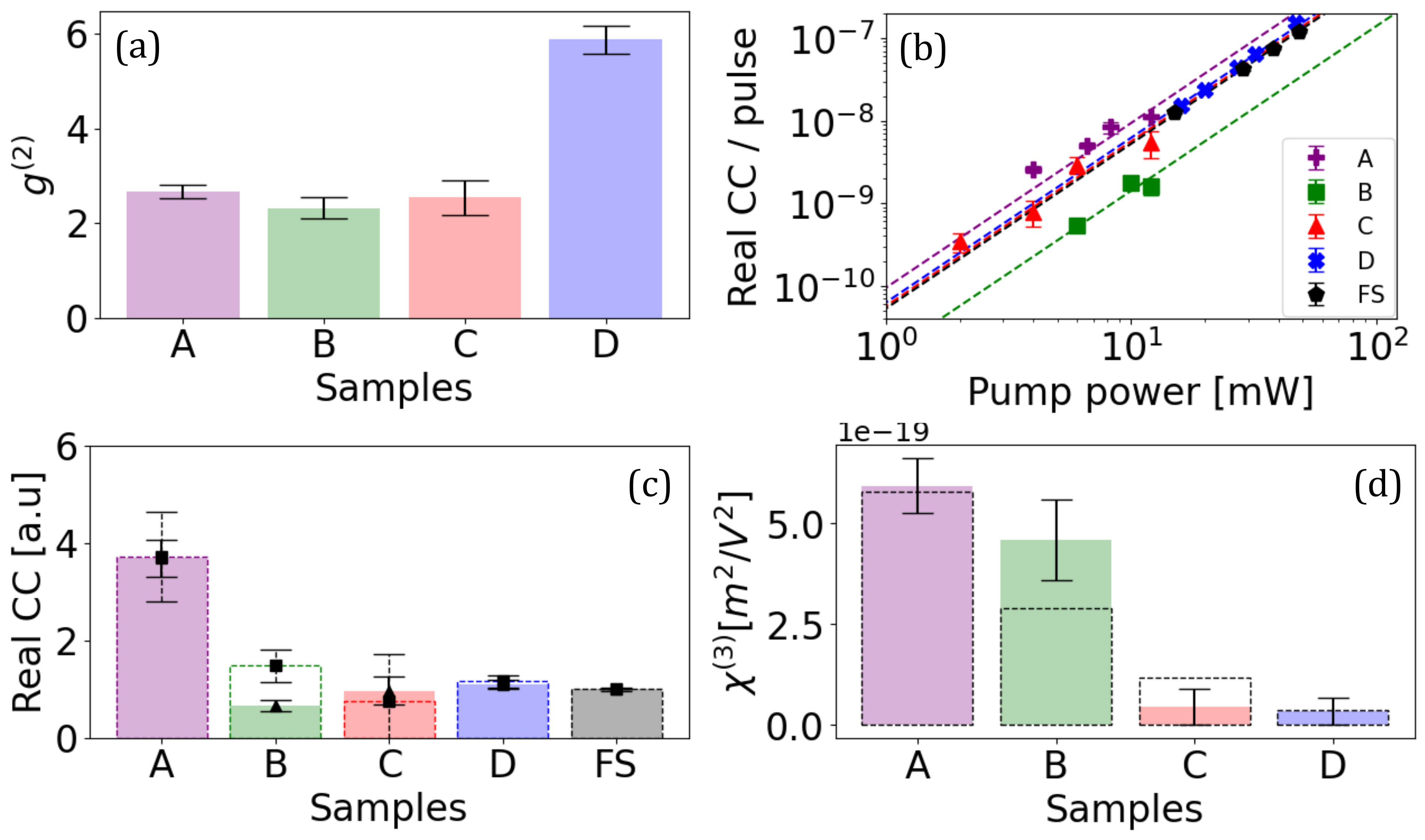}
\caption{(a) Maximal $g^{(2)}$ values obtained for different samples. 
(b) Numbers of real coincidences per pulse vs the pump power for samples A-D and the FS, with quadratic fits. 
(c) Coincidence count rates under the same pump power inside the films for  pairs at 1550 nm, 770 nm (solid) and 1450 nm, 800 nm (dashed). 
(d) The $\chi^{(3)}$ values measured (solid) and estimated from Miller's rule (dashed).}
\label{fig:CC comparison}
\end{figure}

Although single counts are mostly caused by PL, photon pairs can still be identified by registering coincidences between photon detections in signal and idler channels. As expected, the mean number of coincidences per pulse (Fig.~\ref{Data_SiN}c) depends on the pump power quadratically at low powers (solid red line). At stronger pumping, the dependence becomes cubic (dashed green line), due to accidental coincidences resulting from the PL.  
From the coincidence rate, we infer the normalized second-order correlation function at zero delay, defined as $g^{(2)}(0)=\frac{\langle:N^2:\rangle}{\langle\hat{N}\rangle^2}$ and 
 measured as  $g^{(2)}(0)\equiv \frac{\langle N_c\rangle}{\langle N_1\rangle\langle N_2\rangle}$ \cite{ivanova2006multiphoton}, where $\langle N_{c,1,2}\rangle$ are the mean numbers of coincidences and counts in channels 1, 2 per pulse, respectively (Fig.~\ref{Data_SiN} ~(d)). 
The experimental points are well fitted by $g^{(2)}(0)=1+a/P$ \cite{loudon2000quantum}, where $a$ is a constant and $P$ the pump power, because of the linear dependence of the signal in the IR channel on the pump. 
The $g^{(2)}(0)$ dependence on the pump power and its value exceeding 2 confirm the generation of photon pairs. 
Other samples also show high $g^{(2)}(0)$ values, see Fig. \ref{fig:CC comparison} (a). 
The maximal values of $g^{(2)}(0)$ were measured at different pump powers, namely 4, 6, 4, and 16 mW for samples A, B, C, D, respectively, as films with different N concentrations have different damage thresholds. 

In Fig. \ref{fig:CC comparison} (b), we plot the SFWM rates for all samples versus the pump power circulating inside the films (which is different from the incident power due to  etalon effect). 
The mean numbers of coincidences per pulse are plotted after subtracting the numbers of accidental coincidences. 
The latter are estimated as the products of D1 and D2 photocounts per pulse. The scaling, as expected, is quadratic in the pump power for all samples.

The figure also includes the results obtained for the bare fused silica (FS) substrate, which demonstrate similar rates of SFWM. 
This is unsurprising: although $\chi^{(3)}$ for  SiN is three orders of magnitude higher than that of FS ($3\cdot 10^{-22} m^2/V^2$ \cite{boyd2020nonlinear}), the SiN films are three orders of magnitude thinner than the substrate (Table 1).
Note that the SFWM coherence length in FS is so large that almost all the substrate contributes to SFWM.

\section{Two-photon interference}
What could be surprising is that the rate of SFWM from sample B, coated on the substrate, is significantly lower than from the substrate alone. An explanation comes if we account for the quantum interference between the probability amplitudes of a photon pair emission from the SiN film and the substrate. Following the approach of Ref. \cite{klyshko1993ramsey}, we obtain that the two-photon state generated in the ‘film-substrate’ structure is 
\begin{equation}
|\Psi\rangle\propto\bigl[A_f +e^{i(\Delta k_f l+\Delta k_{sub} L)/2}A_{sub}\bigr]|1\rangle_i|1\rangle_s,
    \label{eq: photon state}
\end{equation}
where $A_{f,sub}$ are the amplitudes of pair emission in the film and the substrate, respectively, $\Delta k_{f,sub}$ the corresponding wavevector mismatches, $l$ and $L$  the lengths of the film and the substrate, and $|1\rangle_{s,i}$ signal and idler single-photon states. 
The rate of SFWM pairs is then
\begin{equation}
R\propto A_f^2 +A_{sub}^2 +2 A_f A_{sub} \cos (\Delta\phi),
    \label{eq: rates}
\end{equation}
where the interference phase $\Delta\phi=(\Delta k_{sub}L+\Delta k_fl)/2$. 
Meanwhile, for the substrate alone, the rate is  $R_{sub}\propto A_{sub}^2$.

To calculate $\Delta\phi$, we find that $\Delta k_{sub}L/2=0.51\pi$, while  $\Delta k_lf/2$ are between $0.01\pi$ and $0.08\pi$. Overall, the interference phase is between $0.52 \pi$ and $0.59\pi$, leading to destructive interference for all samples. 
This gives an idea why sample B on a substrate yields fewer pairs than the substrate alone.

The rates of real coincidences under the same pump power inside the films are plotted in Fig.~\ref{fig:CC comparison} (c) as solid bars. 
We now change the interference phase by choosing other wavelengths of the signal and idler photons: 800 nm and 1450 nm, respectively. 
Because of the slow dispersion, the only significant result of passing to these wavelengths is that the phase $\Delta k_{sub}L/2$ becomes $0.39\pi$, which turns destructive interference into constructive one: now, $\Delta\phi$ is from $0.40\pi$ to $0.47\pi$ for different samples. 
The corresponding real coincidence rates (Fig. \ref{fig:CC comparison} (c), dashed bars) are indeed higher than in the previous case, especially for sample B. The reason why only  film B shows  pronounced interference is that its amplitude $A_f$ is close to the one of the substrate. Other films  contribute to SFWM much stronger (sample A) or much weaker (samples C,D) than the substrate. Therefore, they show almost the same pair emission rate for both phases $\Delta\phi$. 

Because the substrate contribution to SFWM is comparable to those of the SiN films, it can be used as a reference to evaluate the third-order susceptibility for all samples. To this end, from Eq.~(\ref{eq: rates}) we find the ratio $A_f/A_{sub}$ for each SiN film. Meanwhile, the amplitude $A_f$ scales as~\cite{wang2001generation}
\begin{equation}
A_f \propto \frac{\chi_f^{(3)}}{n_p} \cdot P_c \cdot l\cdot \mathrm{sinc}(\Delta k_f l/2)\cdot \sqrt{T_s\cdot T_i},
    \label{eq: amplitude}
\end{equation}
where $n_p$ is the pump refractive index, $\chi_f^{(3)}$ the third-order susceptibility for SFWM,  $P_c$  the pump power inside the film, and $T_{s,i}$  the transmission coefficients for the output photons caused by the etalon effect.  
These coefficients vary considerably from sample to sample and with the wavelength; see Supplement 1 Sec.~4 for their calculation and measurement. For the substrate, the etalon effect is negligible because its refractive index is low.

For the substrate contribution, we can similarly write
\begin{equation}
A_{sub} \propto \frac{\chi_{FS}^{(3)}}{n_p} \cdot P_c \cdot L\cdot \mathrm{sinc}(\Delta k_{sub} L/2)\cdot T_p,
    \label{eq:amp_FS}
\end{equation}
where $\chi_{FS}^{(3)}$ is the third-order susceptibility for fused silica and $T_p$ the power transmission coefficient for the pump due to the etalon effect in the film. For SFWM from the substrate alone, there should be no $T_p$ factor. 

Finally, from Eqs. (\ref{eq: amplitude},\ref{eq:amp_FS}), knowing the third-order susceptibility for fused silica, we obtain its values for the SiN films (Fig. \ref{fig:CC comparison} (d)). For samples C and D, the low contribution of the films compared to that of the substrate leads to large uncertainty; as a result, only the upper limit for $\chi^{(3)}$ can be found. The values agree reasonably well with the ones predicted by Miller's rule (dashed bars) but they are considerably lower than for THG because all frequencies involved in SFWM are below the bandgap.  

\section{Conclusion}
In conclusion, we have generated photon pairs through SFWM in sub-wavelength SiN films. Out of the broad SFWM spectrum, we picked two pairs of wavelengths for signal and idler photons by filtering them with band-pass filters.
Although most of the detected photons were from the PL, photon pairs were evidenced through the coincidence measurement. For all samples, values of $g^{(2)}(0)$ exceeded 2 and decreased with the pump power. 
As the nitrogen concentration increases, the $\chi^{(3)}$ reduces and so does the rate of photon pairs generated through SFWM. Additionally, modification of the nitrogen content can dramatically suppress the PL.

Because SFWM has a large coherence length, it inevitably occurs in the substrate as well. 
We show that this can lead to an interference between pair generation in the film and the substrate. 
This fact, not considered so far in the literature, can strongly affect experiments on entangled photon generation via SFWM in ‘flat’ sources. 
In particular, destructive interference can be turned into constructive interference by changing the thicknesses of the film and the substrate or by passing to different wavelengths.

We believe this study demonstrates the possibility of using thin dielectric platforms with third-order nonlinearity for generating entangled photons. 
The pair generation rates in our experiments are very low, and future work will be focused on enhancing the generation rate by structuring the material to obtain nonlinear metasurfaces and using their geometric resonances.

\bibliography{sample}






\end{document}